\documentclass[fleqn,usenatbib]{mnras}

\usepackage{newtxtext,newtxmath}

\usepackage[T1]{fontenc}

\DeclareRobustCommand{\VAN}[3]{#2}
\let\VANthebibliography\thebibliography
\def\thebibliography{\DeclareRobustCommand{\VAN}[3]{##3}\VANthebibliography}

\usepackage{graphicx}	
\usepackage{amsmath}	

\title[Hard State Can't Be MAD]{The Luminous, Hard State Can't Be MAD}

\author[P. C. Fragile et al.]{
P. Chris Fragile,$^{1}$\thanks{E-mail: fragilep@cofc.edu (PCF)}
Koushik Chatterjee,$^{2}$
Adam Ingram,$^{3}$ and
Matthew Middleton$^{4}$
\\
$^{1}$Department of Physics and Astronomy, College of Charleston, Charleston, SC 29424, USA \\
$^{2}$Black Hole Initiative at Harvard University, 20 Garden Street, Cambridge, MA 02138, USA \\
$^{3}$School of Mathematics, Statistics, and Physics, Newcastle University, Newcastle upon Tyne NE1 7RU, UK \\
$^{4}$School of Physics and Astronomy, University of Southampton, Highfield, Southampton, SO17 1BJ, UK
}

\date{Accepted XXX. Received YYY; in original form ZZZ}

\pubyear{2015}

\begin{document}
\label{firstpage}
\pagerange{\pageref{firstpage}--\pageref{lastpage}}
\maketitle

\begin{abstract}
We present a straightforward argument for why the luminous, hard state of black hole X-ray binaries (BHXRBs) cannot always be associated with a magnetically arrested accretion disc (MAD). It relies on three core premises: 1) that the type-C quasi-periodic oscillation (QPO) is best explained by Lense-Thirring (LT) precession of a tilted, inner, hot flow; 2) that observed optical and infrared (IR) QPOs with the same or lower frequency as the type-C QPO suggest the jet, too, must precess in these systems; and 3) that numerical simulations of MADs show that their strong magnetic fields promote alignment of the disc with the black hole and, thereby, suppress LT precession. If all three premises hold true, then, at least whenever the optical and IR QPOs are observed alongside the type-C QPO, these systems cannot be in the MAD state. Extending the argument further, if the type-C QPO is always associated with LT precession, then it would rule out MADs anytime this timing feature is seen, which covers nearly all BHXRBs when they are in the luminous, hard and hard-intermediate states.
\end{abstract}




\section{Introduction}

Black hole X-ray binaries (BHXRBs) generally only reach observable X-ray luminosities whenever they go through an outburst, which typically lasts from a few weeks to many months. These outbursts trace characteristic patterns in hardness-intensity diagrams \citep[HID;][]{Fender04}, with spectral and timing features that are common across many sources \citep{Belloni02,Remillard06}. Because the observational characteristics of these sources evolve in well-defined ways during a given outburst, observers have identified characteristic states that these sources pass through \citep[see][for reviews]{Homan05,McClintock06,Done07}. At the start of all outbursts, sources are found in the ``hard'' state. As the name implies, this state shows a hard spectrum that is often completely dominated by a power-law component. Sources in this state also show steady radio emission, with a flat or inverted spectrum, consistent with a self-absorbed, compact jet \citep{Blandford79}. The jet sometimes also contributes significantly in the infrared \citep[IR;][]{Corbel02} and optical \citep{Russell06} bands. Sources are seen to remain in the hard state up to luminosities of a few tens of percent of Eddington \citep[e.g.][]{Dunn10}, at which point they usually undergo a state transition. Subsequently, they may pass through a hard-intermediate, soft-intermediate, and soft state, before eventually returning to the hard state, although with hysteresis \citep{Miyamoto95,Maccarone03}, and ultimately to quiescence in most cases. Our focus in this work is primarily on the luminous, hard and hard-intermediate states.

Although the exact nature of the hard state is still hotly debated, a picture that has been gaining support in recent years is to associate it with a magnetically arrested disc (MAD) \citep[e.g.][]{Igumenshchev09,Begelman14,Dexter14}. In a MAD, the accreting gas drags in sufficient magnetic flux to saturate the black hole and disrupt the accretion flow \citep{Bisnovatyi76,Narayan03}. This naturally leads to an interruption, or truncation, of the thin disc outside of the normal marginally stable radius. The fast accretion inside the MAD flow prevents efficient radiation, suppressing the soft contribution to the spectrum. Instead, energy may be dissipated via MHD waves or magnetic flares that heat the low-density plasma surrounding this region \citep{Igumenshchev09,Ripperda22}. This hot plasma can then account for the hard, power-law component of the spectrum via Compton scattering of soft photons coming from the truncated, thin disc.

However, this association of the hard state with a MAD stands in direct conflict with a very popular and successful model for the type-C, low-frequency quasi-periodic oscillation (QPO), which is seen in nearly all of these sources when they are in the luminous, hard and hard-intermediate states \citep[see][for a review of QPO phenomenology]{Ingram19}. This model, first described in detail in \citet{Ingram09}, explains the type-C QPO as the result of global, Lense-Thirring (LT) precession of a tilted, inner, hot flow. There are now extensive and diverse observations in support of this general picture, as we summarize in Sec. \ref{sec:disc}.

Furthermore, a related QPO has recently been observed in quasi-simultaneous X-ray, optical \citep{Gandhi10} and infrared \citep{Kalamkar16} observations of multiple sources. The detection of a single QPO frequency across such a wide range of wavelengths strongly points to a common origin. The most likely interpretation is that the optical and infrared signals are coming from precession within the jet itself \citep{Malzac18}, likely in response to precession occurring in the inner accretion flow. The X-ray QPO may come either from the base of the jet or the inner accretion flow \citep{Heil15}. In Sec. \ref{sec:jets}, we review the current evidence for this interpretation.

The conflict arises out of numerical simulation results that, so far, suggest that magnetically arrested discs and their jets do not precess \citep[][Chatterjee et al. in prep]{McKinney13,Ressler23}. This owes to the strong magnetic fields accumulated close to the black hole that are the defining feature of MADs \citep[e.g.][]{Igumenshchev03,McKinney12}. These strong fields force the accretion flow to align with the symmetry plane of the black hole, thereby precluding any LT precession. Since there are still relatively few published simulations of tilted MADs, we recapitulate the latest results in Sec. \ref{sec:MAD}.

If, indeed, MADs cannot precess, then we face one of two inevitabilities: either the luminous, hard state of BHXRBs is not MAD or the type-C and jet QPOs in these sources are not associated with LT precession. In either case, we would be forced to abandon a promising model for understanding BHXRB phenomenology. In Sec. \ref{sec:summary}, we weigh these two possibilities along with the third possibility that current simulations are somehow leading us astray.

\section{Evidence for Precession in Hard State}
\label{sec:disc}

The type-C QPO is the most common QPO seen in BHXRBs. Although there is evidence for it even in soft accretion states \citep{Motta12}, it is most prominent in the luminous, hard and hard-intermediate states. Its frequency is tightly correlated with the outburst cycle, rising from a few mHz near the start of the outburst to $\sim 10$ Hz in the intermediate states. This has now been confirmed in numerous sources \citep{Ingram19}. 

While there are many models for the type-C QPO, the current best observations strongly favor a geometric origin, with precession being the most consistent explanation. Here we summarize those observations and their interpretations:

\begin{itemize}
    \item High-inclination (more edge-on view) sources exhibit stronger QPOs than low-inclination (more face-on view) ones \citep{Motta15,Heil15}. This is directly expected from a precession model, since a precessing corona viewed face on does not modulate the X-ray flux at all, whereas a strong oscillation would be seen from the same corona viewed edge on. 
    \item There is an inclination dependence to the QPO phase lags \citep{vandenEijnden17}. For observations with low QPO frequencies ($\lesssim 2$ Hz), all sources display small ($\sim 0.003$ QPO cycles) hard lags in their QPO fundamental (the maximum count rate in the QPO waveform occurs later in the hard band than in the soft). For observations with higher QPO frequencies, low-inclination sources show large (up to $\sim 0.1$ QPO cycles) hard lags, while high-inclination sources exhibit large (also up to $\sim 0.1$ QPO cycles) soft lags (maximum count rate occurs earlier in the hard band). These results are harder to directly interpret than the amplitude dependence, but are much easier to explain if the QPO is a geometric effect than, say, an oscillation of some intrinsic disc property. Within the precession model, it can be interpreted as a trade-off between two effects. On the one hand, the projected area of the corona is maximal at the point in its precession cycle when it is viewed most face on, while on the other, Doppler boosting is maximal half a precession cycle later when the corona is viewed most edge on (since this maximises the line-of-sight velocities). We may expect projected area effects to dominate for low inclination sources and Doppler effects to dominate for high inclination ones \citep{Veledina13}, leading to high- and low-inclination sources displaying opposite QPO lag behaviour to one another, as observed.
    \item There are also hints that the QPO waveform itself depends on inclination. \citet{deRuiter19} tracked this by measuring the phase difference between the first and second harmonics for many RXTE observations. They found that all high-inclination sources follow a common evolution of phase difference with QPO frequency, whereas low-inclination sources seem to behave independently, again pointing to an inclination dependence of this QPO.
    \item Perhaps the best evidence in favor of the precession model is provided by QPO phase-resolved spectroscopy \citep{Ingram15}. The reflection (or covering) fraction has been found to vary systematically with QPO phase in both H 1743-322 \citep{Ingram17} and GRS 1915+105 \citep{Nathan22}, as has the iron line centroid energy \citep{Ingram16}. Both properties are naturally predicted in the precession model; the former from the angle between the rotation axes of the disc and corona changing as the corona precesses around the black hole spin axis \citep{You20}, and the latter from 
    the corona preferentially illuminating the approaching/receding disc material at different phases of its precession cycle, giving rise to a blue-/red-shifted iron line \citep{Schnittman06,Ingram12}.
\end{itemize}

The sample of sources for all of these studies is relatively small ($\lesssim 15$), but there is high statistical confidence that the QPO amplitude dependence on inclination is real \citep{Motta15}, and detailed statistical analysis concludes that the probability of the inclination dependency for the lags occurring purely by chance is $\approx 0.5\%$ \citep{vandenEijnden17}. In other words, the evidence for the type-C QPO being a geometric effect is quite strong. Additionally, the observed QPO phase-dependence of the iron line centroid effectively rules out most QPO models other than precession. 

As a final note in this section, we acknowledge that evidence for precession is not necessarily evidence for LT precession, which would require the corona be tilted with respect to the symmetry plane of the black hole. The corona could be tilted due to a global tilt of the system (the angular momentum axis of the binary not being aligned with the spin axis of the black hole). GRO J1655-40 \citep{Fragile01}, SAX J1819-2525 \citep{Maccarone02} and MAXI J1820+070 \citep{Poutanen22} all exhibit evidence for this kind of misalignment. Otherwise, it may be that the tilt is somehow self-excited by the accretion flow. 
Whatever the case may be, LT torque is by far the most likely culprit for driving precession over the range of radii and timescales we are talking about here, at least for black holes. Therefore, we take the strong evidence of precession as strong evidence for LT precession.

\section{Evidence for Precession in Jets}
\label{sec:jets}


Direct observation of jet precession is only possible with very long baseline interferometry of discrete ejecta, and even then only for certain active galactic nuclei \citep[AGN; e.g.,][]{Dominik21} and XRBs \citep{Miller-Jones19}. 
In the absence of imaging, the most readily accessible evidence for jet precession may be in the correlated timing properties between spectral bands associated with the disc corona and the body of the jet. 

Observed optical/IR--X-ray lags already suggest a connection between the X-ray corona and the jet \citep[e.g.,][]{Gandhi10,Kalamkar16}. If this is the case, then it stands to reason there may be an optical/IR (i.e., jet) counterpart to any QPO associated with the corona, though this depends on the specific QPO mechanism. For the LT precession model, the jet QPO frequency should be equal to (for rigid precession) or lower than (for differential precession) the type-C QPO seen in X-rays, at least when the source is in the luminous, hard state. 

Thus, we consider correlations between the type-C and optical/IR QPO frequencies to be evidence in further support of precession. However, such correlations are difficult to catch due to the fairly rapid changes in the QPO frequency \citep[on timescales of days or less; e.g.,][]{Hynes03}. The association requires near simultaneous observations across these wavelengths. Here we report on the few sources for which such observations have been made:

\begin{itemize}
    \item {\bf GX 339-4}: During a 1981 outburst of this source, \citet{Motch82} reported QPOs in simultaneous X-ray and optical data. The QPOs were seen to have frequencies of $\sim$0.1 and 0.05 Hz in the X-rays \citep[likely a type-C QPO, given the definitions in][]{Motta11} and $\sim$0.05 Hz in the optical; notably, there was no corresponding optical QPO at $\sim$0.1 Hz. More recently, \citet{Gandhi10} studied a fainter hard state of this source during its 2007 outburst, finding the same optical QPO with a frequency of 0.05 Hz. Although there was no obvious corresponding X-ray QPO in the RXTE data, its presence could not be excluded, and its non-detection may be attributed to the X-ray faintness of the source during this particular outburst \citep{Nowak99}. 
    
    The source was observed again during its 2010 outburst, with QPOs detected in the optical and IR at $\sim$0.08 Hz and an X-ray QPO found at 0.16 Hz, with a weaker signal at $\sim$0.09 Hz \citep{Kalamkar16}. The similarity of the IR/optical QPO frequency to the X-ray signal at $\sim$0.09 Hz led \citet{Kalamkar16} to conclude that the jet was precessing in response to the precession of the hot corona \citep[with the stronger 0.16 Hz signal being a harmonic; see][]{Veledina13}.

    \item {\bf XTE J1118+48}: The rising phase of the 2000 outburst of this source was studied in UV, optical, and X-rays \citep{Hynes03}. QPOs were seen in all three bands with consistent frequencies varying monotonically from 0.08 to 0.13 Hz over a period of 50 days. The X-ray QPO can be identified as type-C according to the definition in \citet{Casella05}.

    \item {\bf MAXI J1820+070}: The 2018 outburst of this source was observed widely, with QPOs reported in X-rays \citep[e.g.,][]{Buisson18,Buisson19} and optical \citep{Mao2022}. The frequencies were around 0.05 Hz in both bands and appear consistent to within $\sim$20\%. The X-ray QPO was consistent with being type-C.  
    



    \item  {\bf BW Cir}: The 2015 outburst of this source showed QPOs in both X-rays and optical at 0.018 Hz \citep{Pahari17}. At the peak of the outburst, the X-ray QPO had increased to $\sim$0.19 Hz, while the optical QPO was not detected. However, this outburst was unusual in that it did not follow the traditional path around the hardness-intensity diagram.  
      
\end{itemize}

Condensing the above, one may tentatively conclude that, during the rising phase of an outburst, where the hard X-ray emission is sufficiently bright, a type-C QPO can be detected with an optical (and maybe UV or IR, where data permits) counterpart at similar frequencies. Given that the IR emission from the disc cannot easily vary on the short timescales of these QPOs, it seems that the best explanation is that the corona and jet are precessing together, producing the respective QPOs \citep{Kalamkar16}.


In addition to the above, precessing jets have been {\it confirmed} during the super-Eddington outburst of V404 Cyg \citep{Miller-Jones19} and the persistently highly super-Eddington source, SS433 \citep{Fabian79, Margon79}. In both of these cases, there is evidence that the disc is inflated \citep[e.g.,][]{Motta17}, which may then precess due to the action of a variety of torques, such as a slaved disc in the case of SS433 (\citealt{Heuvel1980}) and LT in the case of V404 Cyg \citep{Middleton18, Middleton19, Miller-Jones19}. The accretion flow in these systems may well share similarities with the corona in the luminous, hard state (the key difference being the density and mass loss at super-Eddington rates), supporting the general picture of coordinated precession between a thick disc and jet.


\section{No Precession in MAD Simulations}
\label{sec:MAD}

In light of the strong evidence that the inner disc/corona and jets of BHXRBs in the hard and hard-intermediate states precess, we now turn to numerical simulations to see what they have to say. First, in the case of weak magnetization (non-MAD states), such discs {\em do} precess in a manner consistent with type-C QPO behavior \citep{Fragile07,Liska18,White19}, even in the presence of a surrounding thin disc \citep[i.e., in a truncated disc geometry;][]{Bollimpalli23}. Furthermore, the corresponding jets precess {\em in phase} with the disc \citep{Liska23, Ressler23}. In other words, the jet aligns with and follows the angular momentum axis of the disc, not the black hole spin axis, for non-MAD flows, strongly supporting the picture of the type-C and jet QPOs being driven by LT precession. 

\begin{figure}
\includegraphics[width=0.495\textwidth]{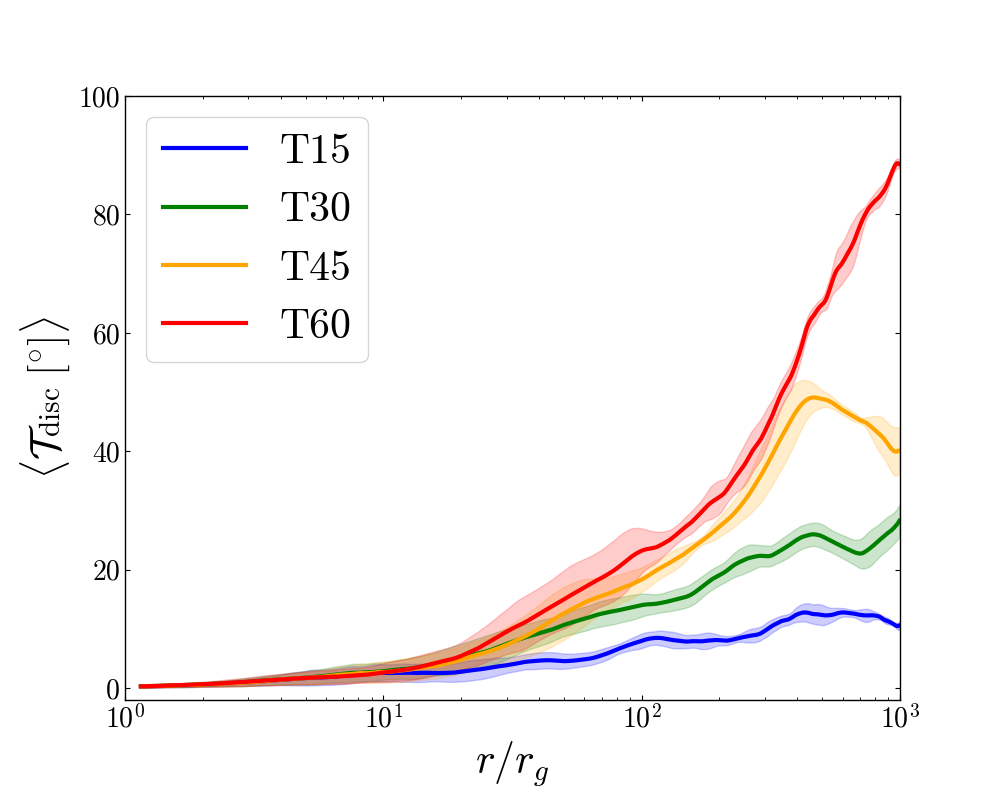}
\caption{Time-averaged (over $20,000 GM/c^3$) radial profiles of the disc tilt angle, $\mathcal{T}_\mathrm{disc}$, with 1$\sigma$ statistical fluctuations for models T15, T30, T45, and T60 of Chatterjee et al., in prep., which had initial tilt angles of 15, 30, 45, and $60^\circ$, respectively. All had $a/M = 0.9375$. In all cases, the inner disc aligns with the symmetry plane of the black hole out to $r \gtrsim 10 GM/c^2$.}
\label{fig:tilt}
\end{figure}

\begin{figure}
\includegraphics[width=0.495\textwidth]{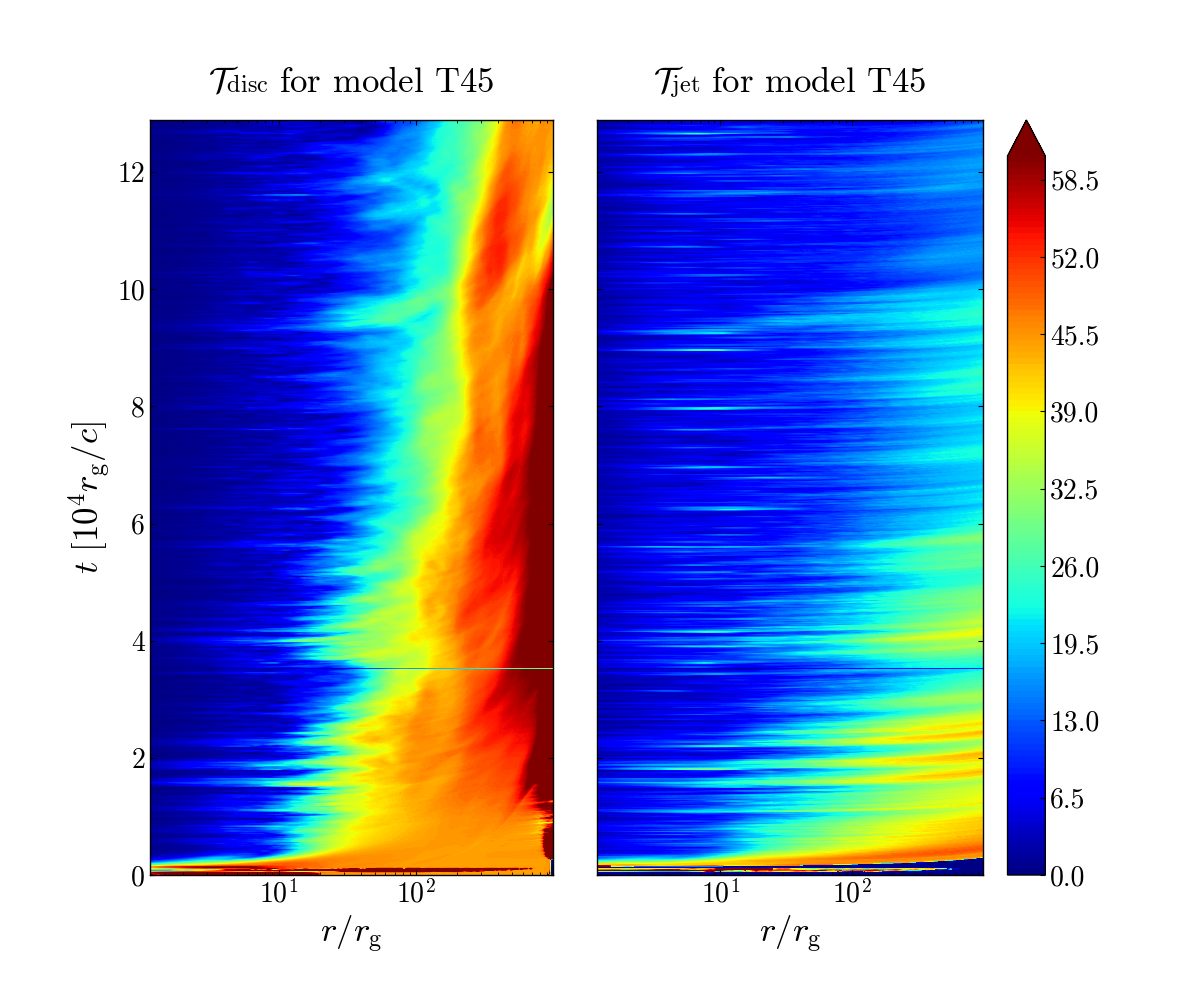}
\caption{Spacetime plots of the disc (left panel) and jet (right panel) tilt angles (in degrees) for model T45 of Chatterjee et al., in prep. As already shown in Fig. \ref{fig:tilt}, the disc quickly aligns (tilt goes to $\approx$0) to beyond a radius of $10 GM/c^2$, but we now also see that the jet direction is nearly perfectly aligned with the black hole spin axis to beyond $10^2 GM/c^2$.}
\label{fig:precession}
\end{figure}

However, for MADs, the case is very different. Here, there is ample and growing evidence that the disc and jet {\em do not} precess \citep[][Chatterjee et al., in prep.]{McKinney13, Ressler23}. This is because, in the MAD state, the black hole ergosphere and inner accretion disc are saturated with magnetic flux. This strong magnetic flux, twisted as it is by the rotating black hole, is forced to align with the black hole spin axis. Because the dynamics of the inner disc are dominated by the magnetic field in the MAD state, the inner disc aligns in response to the alignment of the field. This is demonstrated for a sampling of MAD simulations in Figure \ref{fig:tilt}; please also see Table 1 of \citet{McKinney13}  and Fig. 14 of \citet{Ressler23} for further evidence. We see that even for initial tilts up to $\mathcal{T}_\mathrm{disc} \lesssim 60^\circ$, the inner disc is forced to align with the symmetry plane of the black hole out to $r \gtrsim 10 GM/c^2$. Furthermore, the jet in these simulations also aligns with the spin axis of the black hole, as shown in Figure \ref{fig:precession}. 

 The takeaway message is that MADs force an alignment of the inner accretion flow with the black hole spin axis. This forced alignment prevents LT precession of the disc, which, in turn, precludes any LT-driven precession of the jet. (Because the tilt is effectively $0^\circ$ for much of the jet, it is not even possible to define a reasonable precession angle.)

\section{Summary}
\label{sec:summary}

If everything we have claimed up to this point is true, then we are led to the inevitable conclusion that the luminous, hard state of BHXRBs cannot be MAD. To summarize:
\begin{itemize}
    \item MADs apparently do not precess.
    \item Yet, the type-C and jet QPOs are best explained by precession.
    \item Since the type-C QPO is a ubiquitous feature of the luminous, hard state, it follows that this state cannot be MAD.
\end{itemize}

To some, this may be a dissatisfying conclusion, as there are compelling arguments that the hard state {\em is} MAD. First, the net magnetic flux threading a MAD will increase the strength of the magneto-rotational instability, leading to a high effective viscosity \citep{Begelman14}. This high effective viscosity will prevent the disc from cooling efficiently, leaving a hot, highly magnetized, optically thin flow, as required to produce the hard X-ray spectrum associated with this state \citep{Igumenshchev09,Dexter21}. It is not clear that any other known solution of general relativistic MHD can achieve this. MADs also produce strong jets, which may explain the association of the hard state with persistent, powerful radio emission \citep{Fender01}. Finally, a global magnetic field inversion, in other words, a canceling out of the MAD field, can produce much of the phenomenology associated with the hard-to-soft state transition in BHXRBs \citep{Dexter14}.

So what can we conclude if the hard state is, in fact, MAD? In this case, we may need to confront the possibility that the type-C and jet QPOs are not associated with LT precession, despite the strong supporting evidence provided in Sec. \ref{sec:disc} and \ref{sec:jets}. There are, of course, alternative explanations for the type-C QPO, such as the accretion-ejection instability model \citep[AEI;][]{Tagger99} and JED-SAD picture \citep{Marcel20} \citep[see][for a more complete review]{Ingram19}. However, current iterations of those models do not provide complete pictures of how their respective QPOs operate, nor how they can explain the full set of observations. 

The other alternative is that present simulations are somehow insufficient to support our conclusion that MADs do not precess. Since simulations of tilted accretion discs, whether in the MAD or weakly magnetized states, are still relatively rare, it could be that there are unexplored regions of parameter space where the disc exhibits characteristics of being MAD, yet still retains its tilt and undergoes LT precession. Just to expand on this point, nearly all published tilted disk simulations have assumed dimensionless spins, $a/M \ge 0.9$ \citep[e.g.][]{Liska18,White19,Ressler23}. It could be that at lower spins the torque of the magnetic fields is insufficient to align the disc (evidence to this effect will be presented in Chatterjee et al., in prep.). However, at those lower spins, the torque driving LT precession will also be weaker. 
How all of these effects play out may not be resolved until tilted disc simulations become more commonplace.

As a final note, we call attention to the fact that there is some evidence from Event Horizon Telescope (EHT) observations and matching simulations that both M87 \citep{M87} and Sgr A* \citep{SgrA} host MADs. If we accept this association, then it suggests that MADs may be a common feature of black hole accretion at very low mass accretion rates (corresponding to the quiescent state of BHXRBs). If that is the case, then associating the type-C and jet QPOs with tilt and precession during the luminous, hard state may require that discs transition from a MAD to an un-MAD configuration near the start of their outbursts.

\section*{Acknowledgements}

Thank you to Omer Blaes for the push to write this article and the Lorentz Center, Leiden, The Netherlands, for hosting the ``Overcoming Fundamental Disconnects in our Understanding of Accreting Black Holes'' workshop, where this paper first really began to take shape. PCF is supported by the National Science Foundation through grant AST-1907850. KC is supported by the Black Hole Initiative at Harvard University, which is funded by grants from the Gordon and Betty Moore Foundation, John Templeton Foundation and the Black Hole PIRE program (NSF grant OISE-1743747). AI acknowledges support from the Royal Society. MM is supported by STFC through grant ST/V001000/1.

\section*{Data Availability}

This project did not generate any substantially new data. Data from referenced works are subject to the availability specified in the original publications.



\bibliographystyle{mnras}








\bsp	
\label{lastpage}
\end{document}